\RequirePackage{fix-cm}
\documentclass[smallextended]{svjour3}       
\smartqed  
\usepackage{graphicx}
\usepackage{listings}
\usepackage{xcolor}
\usepackage{subcaption}
\usepackage{longtable}
\usepackage[toc,page]{appendix}
\usepackage{algorithm}
\usepackage{algpseudocode}
\usepackage{multirow}
\usepackage{amsmath}
\usepackage{booktabs} 
\usepackage[most]{tcolorbox}\usepackage{enumitem}
\definecolor{javared}{rgb}{0.6,0,0} 
\definecolor{javagreen}{rgb}{0.25,0.54,0.35} 
\definecolor{javapurple}{rgb}{0.5,0,0.35} 
\definecolor{specialtokenblue}{rgb}{0.25,0.35,0.75} 
\definecolor{backcolour}{rgb}{0.97,0.97,0.96}

\newtcolorbox{conclusionBox}[2][]{%
  attach boxed title to top center
               = {yshift=-8pt},
  colback      = white,
  colframe     = black,
  fonttitle    = \bfseries,
  colbacktitle = white,
  coltitle     = black,
  title        =#2,#1,
  enhanced,
}

\begin{document}

\title{Code Linting Using Language Models
}


\author{Darren Holden        \and
        Nafiseh Kahani  
}


\institute{Darren Holden\at
              Carleton University \\
              \email{darren.holden@carleton.ca}           
           \and
           Nafiseh Kahani \at
              Carleton University\\
              \email{kahani@sce.carleton.ca}  
}

\date{Received: date / Accepted: date}

\maketitle

\begin{abstract}
Code linters play a crucial role in developing high-quality software systems by detecting potential problems (e.g., memory leaks) in the source code of systems. Despite their benefits, code linters are often language-specific, focused on certain types of issues, and prone to false positives in the interest of speed. This paper investigates whether large language models  can be used to develop a more versatile code linter. Such a linter is expected to be language-independent, cover a variety of issue types, and maintain high speed. To achieve this, we collected a large dataset of code snippets and their associated issues. We then selected a language model and trained two classifiers based on the collected datasets. The first is a binary classifier that detects if the code has issues, and the second is a multi-label classifier that identifies the types of issues. Through extensive experimental studies, we demonstrated that the developed large language model-based linter can achieve an accuracy of 84.9\% for the binary classifier and 83.6\% for the multi-label classifier.

\end{abstract}

\lstdefinelanguage{ocl1}{
  keywords={init,Message, Pre, Post, Client,Range,Invariant,Constraints},
  keywordstyle=\color{purple}\ttfamily\bfseries,
  keywords=[2]{not, and, or, state, transition, implies, component,dbg,receipt,:,|,;},
  keywordstyle=[2]\color{black}\ttfamily\bfseries,
  sensitive=false,
  comment=[l]{//},
  morecomment=[s]{/*}{*/},
  commentstyle=\color{blue}\ttfamily,
  stringstyle=\color{red}\ttfamily,
  morestring=[b]',
  morestring=[b]"
}

\lstdefinelanguage{oc}{
  keywords={container, key,list,leaf,type},
  keywordstyle=\color{blue}\ttfamily\bfseries\footnotesize,
  keywordstyle=[2]\color{black}\ttfamily\bfseries\footnotesize,
  otherkeywords = {-,|},
  morekeywords = [3]{-},
  morekeywords = [4]{|},
  keywordstyle = [3]{\color{blue}}\ttfamily\footnotesize,
  keywordstyle = [4]{\color{blue}}\ttfamily\footnotesize,
  identifierstyle=\color{black}\ttfamily\footnotesize,
  sensitive=true,
  comment=[l]{//},
  morecomment=[s]{/*}{*/},
  commentstyle=\color{green}\ttfamily\footnotesize,
  stringstyle=\color{black}\ttfamily\footnotesize,
  morestring=[b]',
  morestring=[b]",
  alsodigit={:}
}

\lstdefinelanguage{octree}{
  keywords={rw, ro},
  keywordstyle=\color{blue}\ttfamily\bfseries\footnotesize,
  keywordstyle=[2]\color{black}\ttfamily\bfseries\footnotesize,
  otherkeywords = {-,|},
  morekeywords = [3]{-},
  morekeywords = [4]{|},
  keywordstyle = [3]{\color{blue}}\ttfamily\footnotesize,
  keywordstyle = [4]{\color{blue}}\ttfamily\footnotesize,
  identifierstyle=\color{black}\ttfamily\footnotesize,
  sensitive=true,
  comment=[l]{//},
  morecomment=[s]{/*}{*/},
  commentstyle=\color{green}\ttfamily\footnotesize,
  stringstyle=\color{black}\ttfamily\footnotesize,
  morestring=[b]',
  morestring=[b]",
  alsodigit={:}
}

\lstdefinelanguage{z3}{
	sensitive=true,
	alsoletter={\-},
	comment=[l]{;},
	keywords=[1]{
apply, assert, assert-soft, check-sat, check-sat-using, compute-interpolant,
declare-const, declare-datatypes, declare-fun, declare-map, declare-rel,
declare-sort, declare-tactic, define-sort, display, echo, eval, exit,
fixedpoint-pop, fixedpoint-push, get-assertions, get-assignment, get-info, get-
interpolant, get-model, get-option, get-proof, get-unsat-core, get-user-tactics,
get-value, help, help-tactic, labels, maximize, minimize, pop, push, query,
reset, rule, set-info, set-logic, set-option, simplify
	},
	morekeywords=[2]{
check-sat-using, declare-var, declare-rel, rule, query, set-predicate-
representation, maximize, minimize, assert-soft, assert-weighted, compute-
interpolant
	},
}
\newcommand{\oc}{\textit{OC }}
\section{Introduction}
Ensuring code quality and maintaining consistent coding standards are essential for creating robust, maintainable, and error-free software systems. Code linting tools, often integrated into Continuous Integration (CI) pipelines, play a crucial role in this process by analyzing source code for potential problems, stylistic inconsistencies, and other issues. However, despite their widespread use, developing efficient and accurate linters poses significant challenges due to several limitations as discussed below.

\noindent \textbf{Language and Domain Specific.}
Most existing linters are designed for specific programming languages. For instance, some tools may support only Java (such as SpotBugs~\cite{spotBugs}), while others may be tailored for C/C++ (such as Cppcheck~\cite{Marjamäki_2007}). Additionally, these tools often focus on particular types of issues rather than providing comprehensive coverage. For example, one linter might specialize in detecting syntax errors and style issues, while another might focus on performance-related problems or security vulnerabilities (e.g. Checkstyle focuses on checking for violations of Java code styling conventions~\cite{checkstyle}).

\noindent \textbf{Accuracy and Performance.} Existing linters rely on source code analysis, which can be categorized into \textit{dynamic} and \textit{static} techniques. Dynamic analysis, while thorough, is often expensive and complicated to implement. Conversely, static analysis tends to overestimate issues and suffers from accuracy problems. This creates a situation where selecting the appropriate type of linter involves a trade-off between performance, computational resources, and accuracy. This issue is particularly critical in the context of CI, where fast CI cycles are crucial for effective and agile software development practices.

\noindent \textbf{Evolving Coding Standards, Issues and Practices.} 
Programming practices and issue types, especially security and vulnerability issues, evolve constantly. Therefore, keeping tools updated to detect new types of issues and comply with new standards and practices poses significant challenges for maintaining effective linters.

Large language models such as OpenAI's GPT-4 \cite{achiam2023gpt}, have demonstrated  proficiency in understanding and generating human language, including programming languages. These models are trained on diverse datasets encompassing vast amounts of textual and code-based information, enabling them to capture complex patterns, semantics, and contextual nuances in code. We envision that leveraging language models  for code linting can eventually address the aforementioned challenges. More specifically:

\noindent \textbf{Language Agnostic}. Language models can be trained on multiple programming languages, making them versatile and capable of analyzing diverse codebases. This adaptability reduces the need for multiple language-specific linters tailored to different programming languages and issue types. Additionally, unifying efforts on developing a single, comprehensive linter that supports all languages can lead to more focused and efficient endeavors, ultimately creating a more robust and effective tool.

\noindent \textbf{Improved Accuracy and Performance.} Language models can analyze code within its broader context, considering not just syntax but also the logical flow and intent behind the code. This contextual understanding allows for more precise identification of potential issues and more meaningful suggestions for improvement. Although training language models can be time-consuming, their predictions are fast and can even surpass the speed of static analysis techniques.

\noindent \textbf{Continuous Learning and Updates.} By continually training on new code repositories and integrating feedback from developers, language models can evolve to stay current with emerging coding trends and practices. This ensures that the linting process remains relevant and effective, capable of detecting new types of issues and adhering to updated standards.

Despite extensive research over the last few years on program generation, repair, and code comprehension, there has been relatively little research directly addressing the use of language models for code linting. This paper explores the architecture and implementation of a language model-based code linter and evaluates its performance under different configurations. To achieve this, we first create a large dataset of code snippets and their corresponding issues by analyzing open repositories. We then develop two classifiers using an existing Code Language Model (CLM): one to detect if a given code snippet contains an issue, and another to predict the type of issue. Relying on the collected dataset, we answer the following Research Questions (RQs).

\begin{itemize}
    \item \textbf{RQ1} \textit{How does the selected CLM perform in detecting code issues?} We find that our approach is able to achieve an accuracy of 84.0\% in binary issue detection, and an F1 score of 0.838 in issue identification. It is also able to correctly detect a higher percentage of issues than is reported by the static analysis linters while being 67.92\% faster than the applied linters on average.
    
    \item \textbf{RQ2} \textit{How do different input formulations affect the model's performance?} We find that by removing certain parts of the input code (such as comments, or Javadoc) can have an impact on the performance of the model. Certain input formats can result in accuracy improvements of up to 3.1 percentage points. 
    
    \item \textbf{RQ3} \textit{How does the fine-tuned model perform when identifying issues of types that are rare in the collected dataset?} Our analysis shows that there is a significant performance decrease (an accuracy decrease of 31.1 percentage points) in issue detection when a CLM-based linter encounters issues that were not present in the training dataset. Similarly, in issue identification, the model achieves a F1 score of 0.912 on the issue types that are commonly found in the dataset, and a score of 0.644 of the issue types that are sparsely found in the dataset.
    
    \item \textbf{RQ4} \textit{How does our model's performance differ when analyzing projects that are present in the pre-training dataset compared to those that are not?} The results show that there is a significant improvement in our model's performance when analyzing projects that were present in the pre-training dataset, regardless of if those projects were included the fine-tuning dataset. The model achieves accuracy that is up to 11.6 percentage points higher on projects that were in the pre-training dataset compared to those that were not.
\end{itemize}

The rest of this paper is organized as follows. Section \ref{ch:background} introduces the foundational concepts on which our work is based and examines existing studies relevant to our work. Section \ref{ch:approach} describes our approach in detail. Section \ref{ch:evaluation} presents our approach for linter tool and model selection. It also presents the results of our experimental study, addressing four key research questions.

\section{Background}\label{ch:background}
This section  covers some background concepts related to this study, and  provides a brief overview of  related work in this area. 

\subsection{Code Linting}
Code linters are tools  designed to automatically detect code issues  by analyzing code, helping   developers improve their code's readability, maintainability, and security~\cite{vayadande2023let}.
Linters can detect a wide variety of issues, including defects, code styling issues, and design issues~\cite{vassallo2020developers}. They can also enforce coding styles~\cite{johnson2013don}, automate code reviews~\cite{vassallo2020developers}, and measure code metrics (such as cyclomatic complexity)~\cite{vassallo2020developers}. Code linters can utilize both static and dynamic analysis techniques in order to detect code issues. Dynamic analysis is the less common approach, but involves executing the code to observe its behaviour~\cite{gong2015dlint}, while  static analysis, which is the focus of this project, finds issues by analyzing the source code without execution~\cite{emanuelsson2008comparative}. Static analysis linters include  data-flow analysis and control-flow analysis~\cite{beller2016analyzing}, as well as more abstract techniques such as pattern matching, or bug-specific heuristic patterns~\cite{ayewah2008using}. Some examples of available code linting tools include SpotBugs~\cite{spotBugs}, Infer~\cite{infer}, SonarCloud~\cite{SonarCloud}, CppCheck~\cite{Marjamäki_2007}, and Coverity~\cite{Coverity}.

A common problem with linters is their tendency to produce \textit{false positive} reports (cases where a reported issue with the code is not an actual issue), which can be more numerous than true positive reports in some cases~\cite{johnson2013don}. This problem arises from the undecidable nature of many software traits that static analysis linters check for, in that it can be infeasible to determine if the trait actually exists in the software or not~\cite{emanuelsson2008comparative}. Due to the undecidable nature of this problem, many linters aim to report no false negatives (cases where an issue with the software is not reported), while minimizing  the number of reported false positives~\cite{emanuelsson2008comparative}. Despite the drawback of reporting many false positives, linters are often used in the software development process, with a significant proportion using more than one tool~\cite{beller2016analyzing}. This is due to their numerous benefits, which extend beyond just finding issues. 

\subsection{Neural Language Models}
This study  uses Neural Language Models (NLMs) to analyze code and predict if that code has issues. NLMs leverage neural networks to learn the features of text data, which are then used for  tasks such as  text summarization, text classification, translation, and question answering~\cite{zhao2023survey}. In  this study, the task of interest is text classification, specifically classifying source code as having issues or not.

The NLM used in this study uses the sequence-to-sequence Transformer architecture~\cite{feng2020codebert}. A Transformer model takes in a text input as a sequence of tokens, encodes it into a continuous representation, and then decodes that representation into the output representation~\cite{vaswani2017attention}.  Transformer models  use  self-attention   to capture dependencies and relationships between the input sequence's tokens~\cite{vaswani2017attention}. The Transformer architecture consists of an encoder followed by a decoder, each with a sequence of layers. Each encoder layer consists of two sub-layers (a multi-head self-attention mechanism, and a fully connected feed-forward network), while each decoder layer consists of three sub-layers (a masked multi-head attention mechanism, a multi-head self-attention mechanism, and a fully connected feed-forward network)~\cite{vaswani2017attention}. These combinations of layers allows the model to generate and then utilize multiple linear transformations of the input sequence, enabling it to observe the entire input sequence and the incomplete output sequence~\cite{vaswani2017attention}.

Furthermore, NLMs are typically pre-trained on a task or set of tasks  to learn general semantic features. These pre-trained models can then be fine-tuned to more specific downstream tasks, generally resulting in improved performance~\cite{zhao2023survey}. Pre-training tasks can include tasks such as masked language modelling, denoising, sentence prediction, and sentence order prediction~\cite{zhou2023comprehensive}. This study focuses on models which are pre-trained on a code-based dataset for code-related tasks, as opposed to models that are solely pre-trained for natural language tasks. 

\subsection{Related Work}
There has been a lot of interest in issue detection in recent years, with many different approaches being studied. This study focuses on utilizing a Transformer-based model for issue detection. 
However, there are many other approaches that utilize neural~\cite{nguyen2022regvd}, \cite{li2021vulnerability}, \cite{chakraborty2021deep}, \cite{li2021vuldeelocator} and non-neural~\cite{pearce2022asleep}, \cite{bian2018nar} methods.

Gao et al. analyzed the performance of 16 different language models on vulnerability detection, a specialized form of issue detection focused on detecting security vulnerabilities~\cite{gao2023far}. They found that GPT models performed  best in both binary classification and multi-class classification tasks. Interestingly, they found that models with higher numbers of parameters did not necessarily result in significant performance improvements. Similarly, Yuan et al. compared 10 different language models on several software development tasks, including issue detection~\cite{yuan2023evaluating}. They focused on a comparison of zero-shot, few-shot, and fine-tuning methodologies. They found that the best model after fine-tuning achieves 61.0\% accuracy on issue detection, while the best accuracy amongst the zero-shot and few-shot methodologies was 55.4\% accuracy. 

Zhou et al. studied how unbalanced datasets  impact the performance of language models in vulnerability detection, amongst other software development tasks~\cite{zhou2023devil}. They found that the studied models are  at least 39\% more accurate in vulnerability detection when identifying ``head'' labels (the commonly occurring labels) than ``tail'' labels. For the best performing model, they reported an overall accuracy of 73.1\%, with  87.0\% accuracy for head labels and  60.6\% for  tail labels. This study highlights the importance of a balanced dataset in this field, and proposes analyzing head and tail labels separately. Fu and Tantithamthavorn proposed LineVul for vulnerability detection~\cite{fu2022linevul}, achieving 
F1, precision, and recall scores of 0.91, 0.97, and 0.86 respectively, when predicting if a function has a vulnerability, and an accuracy of 0.65 for determining the line of code with the vulnerability.

On top of specific approaches, there have been quite a few surveys done reviewing the current state of issue detection, often with a specific focus on vulnerability detection. Recently, Nong et al. performed a review of how the practice of open science is handled amongst studies which apply deep learning to vulnerability detection~\cite{nong2022open}. They analyzed the accessibility, executability, reproducibility, and replicability of many studies,  finding that each of these areas is currently lacking in the field. Another survey performed by Bi et al. looks at benchmarking approaches for vulnerability detection, and the lack of a comprehensive approach~\cite{bi2023benchmarking}. Some of the challenges they highlight is the lack of availability in datasets (resembling the lack of availability noted by Nong et al.~\cite{nong2022open}), a lack of a commonly used evaluation methodology, and a lack of a sufficiently large, accessible dataset. Steenhoek et al. compared many different vulnerability detection approaches on two datasets~\cite{steenhoek2023empirical}. They found that many of the approaches performed similarly to each other, although the types of vulnerabilities that each model performed best on varied. Additionally, they found that the different approaches often have an overlap in  lines of code they find most important when identifying vulnerabilities. Finally, Harzevili et al. present a comprehensive overview of machine learning approaches for vulnerability detection~\cite{harzevili2023survey}. Their overview includes the types of models used, the characteristics of the utilized datasets, and the challenges facing studies in this area.

A common gap in this field is that existing work has a significant focus on vulnerability detection rather than general issue detection. Our work aims to address this gap, by evaluating how a language model performs when identifying a wide variety of issues in source code. While existing work focuses solely on detecting vulnerabilities~\cite{gao2023far}, \cite{zhou2023devil}, \cite{yuan2023evaluating}, and \cite{fu2022linevul}, our approach aims to be a general purpose linter, identifying issues related to performance, code organization, and code clarity, on top of issues related to potential vulnerabilities. In terms of utilized datasets, existing studies use datasets with sizes ranging from 108 samples~\cite{gao2023far} to 188,636 samples~\cite{fu2022linevul}, and have varying numbers of labels ranging from binary classification of 2 labels~\cite{yuan2023evaluating} to 113 labels~\cite{zhou2023devil}. In comparison our dataset contains 169,494 samples with 118 labels.

\section{Approach}\label{ch:approach}
\begin{figure}[!t]
    \centering
    \includegraphics[width=0.85\columnwidth]{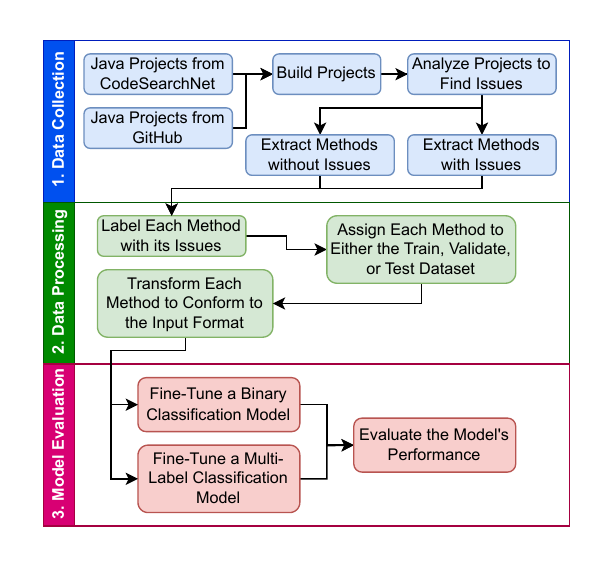}
    \caption{Overview of the approach utilized in this study}
    \label{fig:approach}
\end{figure}

The objective of this study is to determine if a language model can identify issues in a given code-base. Specifically, we aim to explore whether a language model can be tuned to perform code linting, a task that typically requires computationally expensive analysis. To achieve this goal, our approach follows several steps, as illustrated in Figure~\ref{fig:approach}: 
\begin{itemize}
    \item \textbf{Data Collection}. Methods with and without issues are collected, as discussed in Section \ref{Target-Methods}.
    \item \textbf{Pre-processing of the Collected Data}. The collected methods are appropriately formatted for analysis.
    \item \textbf{Tuning and Application of the Language Model}. A language model is tuned and applied to the formatted methods to identify potential issues.
\end{itemize}


\subsection{Data Collection}\label{Target-Methods}
The approach focuses on analysing and detecting issues at the method-level. Thus, the data collection involves extracting source code of methods, referred to as the \textit{target methods} in the rest of this paper. Each target method includes the entire method's \textit{body}, \textit{signature}, any associated \textit{annotations}, and any \textit{comments and Javadoc}. No further context--such as the class containing the target method, or any methods that the target method may call--is provided to the model. Figure~\ref{fig:sampleTargetMethod} illustrates a simple target method example. Each target method is associated with any issues  detected within the method (or lack thereof). For instance, the target method illustrated in Figure~\ref{fig:sampleTargetMethod} would have an associated issue detailing the potential for a \textit{NullPointerException} to be thrown. 

\begin{figure}[htbp]
    \centering
    \begin{lstlisting}[language=Java,numbers=left,breaklines=true]
    /**
    <@\quad@>* Prints an attribute of an Object to the console
    <@\quad@>* @param input The object whose attribute should be printed
    <@\quad@>*/
    @Override
    public void printAttribute(SomeObject input) {
    <@\quad@><@\quad@>// Print the attribute
    <@\quad@><@\quad@>System.out.println("Attribute Value: " + input.getAttribute().toString());
    }
    \end{lstlisting}
    \caption{A sample target method with a potential NullPointerException (as the input object may be null)}
    \label{fig:sampleTargetMethod}
\end{figure}

Maintaining minimal context for the target methods offers several advantages. Firstly, this minimal context allows the approach to be applied as soon as the target method is complete, potentially even before any of the target method's dependencies  exist. 
This means that linting can be applied earlier in the development process, enabling potential issues to be addressed sooner. Secondly, minimal context  allows for quicker analysis. By only requiring a string representation of the target method, any overhead due to compiling the source code is avoided, and only the uncompiled target method needs to be extracted. Lastly, using a minimal context decreases the size of the input to the language model. This reduction enables the use of language models with smaller maximum input lengths, making the approach accessible and requiring minimal resources.

\subsection{Pre-processing} \label{inputSection}
Three operations are defined to modify the target method so that different pieces of information in the  target method can be analyzed for their impact on the model's performance. These operations are designed to preserve the target method's behavior and characteristics, ensuring that the source code is not functionally impacted. The operations are as follows:

\noindent \textbf{Remove Comments}. All comments (both single-line and multi-line) are removed from the target method. An example of the result can be seen in Figure~\ref{fig:no-comments-input}. 

\noindent \textbf{Remove Javadoc}. Any provided Javadoc is removed from the target method. An example of the result can be seen in Figure~\ref{fig:no-javadoc-input}. 

\noindent \textbf{Replace String Literals with a Token}. All literal string values in the target method are replaced with a special token. An example of the result can be seen in Figure~\ref{fig:no-strings-input}. 

Using these three operations, the model is tested on eight different input formats: one consisting of the unmodified target methods (as seen in Figure~\ref{fig:sampleTargetMethod}), and seven comprised of the possible combinations of applying the three operations. The most extreme format involves the application of all three operations, as seen in  Figure~\ref{fig:all-ops-input}. 


\begin{figure*}[htbp]
    \centering    
    \begin{subfigure}{\linewidth}
        \begin{lstlisting}[language=Java,numbers=left,breaklines=true]
    /**
    <@\quad@>* Prints an attribute of an Object to the console
    <@\quad@>* @param input The object whose attribute should be printed
    <@\quad@>*/
    @Override
    public void printAttribute(SomeObject input) {
    <@\quad@><@\quad@>System.out.println("Attribute Value: " + input.getAttribute().toString());
    }
        \end{lstlisting}
        \subcaption{Target Method With Comments Removed}
        \label{fig:no-comments-input}
    \end{subfigure}
    
    \begin{subfigure}{\linewidth}
        \begin{lstlisting}[language=Java,numbers=left,breaklines=true]
        @Override
        public void printAttribute(SomeObject input) {
        <@\quad@><@\quad@>// Print the attribute
        <@\quad@><@\quad@>System.out.println("Attribute Value: " + input.getAttribute().toString());
        }
        \end{lstlisting}
        \subcaption{Target Method With Javadoc Removed}
        \label{fig:no-javadoc-input}
    \end{subfigure}
    
    \begin{subfigure}{\linewidth}
        \begin{lstlisting}[language=Java,numbers=left,breaklines=true]
        /**
        <@\quad@>* Prints an attribute of an Object to the console
        <@\quad@>* @param input The object whose attribute should be printed
        <@\quad@>*/
        @Override
        public void printAttribute(SomeObject input) {
        <@\quad@><@\quad@>// Print the attribute
        <@\quad@><@\quad@>System.out.println(<@\textcolor{specialtokenblue}{[<stringliteral>]}@> + input.getAttribute().toString());
        }
        \end{lstlisting}
        \subcaption{Target Method With String Literal Replaced by Special Token}
        \label{fig:no-strings-input}
    \end{subfigure}
    
    \begin{subfigure}{\linewidth}
        \begin{lstlisting}[language=Java,numbers=left,breaklines=true]
        @Override
        public void printAttribute(SomeObject input) {
        <@\quad@><@\quad@>System.out.println(<@\textcolor{specialtokenblue}{[<stringliteral>]}@> + input.getAttribute().toString());
        }
        \end{lstlisting}
        \subcaption{Target Method With All Operations Applied}
        \label{fig:all-ops-input}
    \end{subfigure}
    
    \caption{Examples of several input formats, applied to the sample target method shown in Figure~\ref{fig:sampleTargetMethod}.}
    \label{fig:input-formats}
\end{figure*}

CodeBERT, the language model utilized for this study, has a maximum input length of 512 tokens~\cite{feng2020codebert}. More than 99\% of the target methods (and thus the input) are shorter than this limit. However, any methods that exceed 512 tokens are too long for CodeBERT's input capacity and are therefore truncated. The target methods are truncated after they have been modified with any applicable modification operations.

\subsection{Model Training}
We trained LLMs on two types of tasks. The first is a \textit{binary classification} task to determine  whether the provided target method has an issue or not. For this task, the model output is  one of two labels. One label denotes the \textit{presence of an issue}, and the other indicates that \textit{no issues were found}. The second task is \textit{multi-label classification}, identifying which issues the target method may have. To perform multi-label classification, the output was formatted as a binary output for each issue type, along with a binary output for a ``no issues'' type. This no issues type was included to ensure the model's output explicitly indicates when no issues are detected.

Our implementation uses the HuggingFace Transformers library~\cite{wolf2020transformers}. We employ a pre-trained model, retrieved from HuggingFace (the selected model is specified in Section~\ref{sec:modelSelection}). We also utilize the \textit{AutoModelForSequenceClassification} from the Transformers library~\cite{automodels}, which loads the pre-trained model, and adds a final fully-connected layer for classification, with one output for each label. Each generated model was then trained on the collected dataset, as described in Sections~\ref{sec:dataset} and \ref{sec:defaultSplits}, using the Trainer class from the Transformers library~\cite{trainer}. 

\section{Evaluation}\label{ch:evaluation}
This section details the execution and evaluation the approach. It also includes a detailed discussion of the results. 

\subsection{Linter Selection}
To conduct this study, we  selected static analysis linters for identifying issues in Java code and creating a dataset. Given the large number of static analysis linters available, several criteria were used to narrow down the possible options. The following section discusses these criteria.  

\noindent \textbf{Availability}. For this study, only linters that were freely available were considered. This decision was made in order to make the data collection methodology more accessible and reproducible. This excluded tools such as Coverity~\cite{Coverity}.

\noindent \textbf{Language Analyzed}. As this study focused on Java code, only linters capable of analyzing Java code were considered. This excluded tools such as Cppcheck, which only analyzes C and C++ code~\cite{Marjamäki_2007}.

\noindent \textbf{Automatable}. To simplify the data collection process, only linters that could be run locally through a terminal command were considered. This excludes tools such as SonarCloud, which is offered as a remotely-hosted Software as a Service (SaaS)~\cite{SonarCloud}.


\noindent \textbf{Issue Location}. To build a coherent dataset, it is necessary to identify the location of the discovered issues. Therefore, only tools that provide this location information were considered. This is a very common feature among static analysis linters, so none of the investigated tools were disqualified as a result of this criterion.

Based on these criteria, two static analysis linters were selected for use in this study: Infer v1.1.0~\cite{infer} and SpotBugs v4.8.3~\cite{spotBugs}.

\subsection{Dataset}\label{sec:dataset}
\subsubsection{Project Selection.}\label{CollectedProjects} A large dataset was constructed by using Infer and SpotBugs to analyze a variety of open-source Java projects. The initial list of projects was sourced from the CodeSearchNet dataset~\cite{husain2020codesearchnet}. This list was then refined based on two criteria: (1) the projects must be publicly available on GitHub, and (2) their build processes must use Maven. The requirement for GitHub availability  ensures that the full project source code is easily accessible. The requirement of using Maven for the build process is due to the fact that the selected static analysis linters require compilation of the source code. Requiring that all considered projects use Maven for their build process allows for uniform automation of the project analysis and ensures that project dependencies are easily accounted for. Each selected project was required to have a Maven pom file located at the root of the project directory. Projects were not excluded based on any other specific characteristics (such as number of GitHub stars, number of forks, and repository age) to avoid potential biases in the dataset. This step resulted in 3,268 candidate projects.

The list of candidate projects was expanded by querying the GitHub REST API~\cite{githubrest} for projects written in Java that are tagged with the Maven topic. Unfortunately, the GitHub API limits queries to only return 1,000 results, which was insufficient for the number of candidate projects we wished to collect. To collect additional data, the query was run repeatedly with a varying term specifying the age of the most recent commit. This allowed us to construct a list of new projects by repeatedly querying the API for older and older projects. This list of projects was then filtered to remove any projects that were included in CodeSearchNet, and  any projects that did not have a root-level pom file (as with the CodeSearchNet projects). Using this methodology, we retrieved  3,209 new candidate projects, bringing the total number of candidate projects to 6,477. 

\subsubsection{Analysis of the Candidate Projects.} Infer and SpotBugs were then used to analyze each of the candidate projects retrieved in Section \ref{CollectedProjects}. Each  tool was configured to collect all non-experimental bugs available in each tool by default. Since these tools require each project to be compiled, Maven  v3.6.3 was used along with Java versions 8, 11, or 17, depending on what was specified in the project's pom file. Due to the number of projects requiring analysis, each analysis was allowed to run for only 25 minutes, after which the process was terminated. If an analysis failed for a reason other than reaching the 25-minute time limit, two steps were taken to maximize the number of discovered issues. First, the compilation was re-attempted with the other Java versions. Attempting to compile with other Java versions accounted for cases where the Java version collected from the pom file was incorrect. Second, the compilation was re-attempted with any other poms in the project. Attempting to compile with other poms allowed for the potential of finding some issues (e.g., by analyzing a sub-module of the project in question) instead of discounting the project entirely. 

Each  discovered issue was then mapped to the Java method it was contained in. The scope of this project is limited to considering Java methods, so any issues  discovered outside the scope of a method were discarded. Using this methodology, 2,907 projects had at least one discovered issue, resulting in a total of 108,182 issues from 84,747 methods. To round out the dataset, methods with no detected issues were randomly selected so that the number of methods with no issues was equal to the number of methods with issues. This resulted in a dataset containing 169,494 total Java methods. The 108,182 found issues are composed of 118 types of issues (10 from Infer, and 108 from SpotBugs). All  the collected issue types are listed and described in Table~\ref{table:issueDefs}. This table also specifies IDs  used to refer to each issue type. IDs beginning with `I' refer to issues reported by Infer, and IDs beginning with `S' refer to issues reported by SpotBugs.

\subsubsection{Equivalent Issue Mapping.}
In order to reduce model confusion, the identified issue types were analyzed to convert equivalent issue types into a single type. Equivalent issues types are defined as issue types which are very similar to each other. This ensures that equivalent issues are all labelled the same way, rather than having multiple labels that essentially denote the same issue. To accurately categorize the issues reported, we first reviewed detailed documentation describing different types of issues~\cite{inferDesc}, \cite{spotBugsDesc}, and then manually reviewed collocated issues. 

The simplest case of equivalent issues involves multiple issue types reported by a single tool having  identical or nearly  identical definitions. For example, issues \textit{I1} and \textit{I2} have the exact same definition according to the Infer documentation~\cite{inferDesc}. 

Following the analysis of the documentation, we manually reviewed random samples from the dataset to ensure that equivalent issues were classified under the equivalent types. The samples were randomly selected such that only methods with multiple reported issues on the same line were reviewed, as it is assumed that equivalent issues would often be collocated. In this way, 400 samples were manually reviewed to obtain a 95\% confidence level, and a 5\% margin of error. Collocated issues were compared to their respective issue types' documentation to determine if they were equivalent. For instance, it was found that issue \textit{I3}, and issue \textit{S6} were often collocated. According to the Infer documentation, the \textit{I3} refers to cases where a resource may not be closed, especially if an exception occurs while accessing the resource~\cite{inferDesc}. In the SpotBugs documentation, the \textit{S6} issue refers to cases where a ``stream, database object, or other resource requiring an explicit cleanup operation'' may not be properly closed~\cite{spotBugsDesc}. Manual review of this information determined that these two types of issues are equivalent. 

There were many cases in which collocated issues were similar enough to be deemed equivalent, but there were also cases in which collocated issues were not similar enough to be considered equivalent. For example, consider issues \textit{I5} and \textit{S8}, which were often collocated. Issue \textit{I5} refers to cases where there is a potential data race~\cite{inferDesc}, while issue \textit{S8} refers to cases where references to an object's mutable field are returned, potentially leading to unchecked and unexpected changes to the object's field~\cite{spotBugsDesc}. Although \textit{S8} could lead to a potential data race, it has a much broader application, so these two issues were not marked as equivalent. The collected groups of equivalent issues can be seen in Table~\ref{table:equivIssues}.

\begin{table}[!htb]
    \centering
    \begin{tabular}{p{0.05\linewidth} p{0.3\linewidth} p{0.6\linewidth}}
        \toprule
        ID & Base Issues & Description \\
        \midrule
        E1 & I1, I2, S1 & Issues which denote a possible null dereference \\ \hline
        E2 & I3, S6, S7 & Issues which denote a case where a resource may not be properly closed \\ \hline
        E3 & I4, S2 & Issues where keySet() is used to iterate over a HashMap instead of entrySet() \\ \hline
        E4 & S36, S107 & Issues where the return value of the read() or skip() InputStream methods is ignored \\ \hline
        E5 & S10, S11 & Issues where reflection is used to increase the accessibility of a method or class \\ \hline
        E6 & S4, S43 & Issues where a method is invoked in an incorrect, inefficient, or insecure way \\ \hline
        E7 & S31, S59 & Issues where '==' or '!=' is being used instead of an equals() method \\ \hline
        E8 & I5, S9, S19, S24, S25, S35, S50, S51, S52, S53 & Issues where the implemented synchronization could result in a race condition \\ \hline
        \bottomrule
    \end{tabular}
    \caption{Equivalent Issues and their Encompassed Base Issues }
    \label{table:equivIssues}
\end{table}

\subsubsection{Dataset Imbalance.}
The collected dataset is very unbalanced in terms of the distribution of issue types. A few types are very numerous  (such as issue type \textit{S8}), while many others have very few occurrences (such as issue type \textit{S61}). To better balance the dataset, the least frequently occurring issues are removed from the dataset during fine-tuning and evaluation, which is only performed with the top 75\% most frequently occurring issues. This includes all issue types that occur at least 15 times. With this filter applied, the dataset contains 84,664 methods with issues, and 75 issue types. Random methods without issues were removed until there were 84,664 remaining in the dataset. This is the case unless otherwise specified, and other values for the issue count cutoff are explored as part of \textbf{RQ3}.

\subsubsection{Dataset Splits.}\label{sec:defaultSplits}
The dataset is  split into three sets: the train set, the validation set, and the test set. This is performed by allocating 80\% (169,328 methods) of the dataset to the training set, and 10\% (16,932 methods) each to the validation and test sets. This approach aligns with other studies in this area~\cite{li2021vulnerability}, \cite{steenhoek2023empirical}. An important note is that each of the three sets was selected such that 50\% of the samples in each set have issues, and the remaining 50\% of the samples do not. 

\subsection{Model Selection}\label{sec:modelSelection}
In this study, we aim to fine-tune a language model capable of reproducing the results of static analysis linters. Therefore, the language model was chosen with care as it is an integral part of this study. We used the following criteria to narrow down the options from the many available language models. 

\noindent \textbf{Open Source}. To facilitate fine-tuning and replicability, this study focused on models which are open-source. This excludes models such as GPT-4~\cite{achiam2023gpt}.

\noindent \textbf{Pre-Trained on Code}. We focused on language models that were pre-trained on a corpus including code. The problem of identifying issues in code relies on an understanding of code semantics and structure, so having a model pre-trained on code-related tasks is advantageous. This excludes language models such as T5~\cite{raffel2020exploring}, which is only pre-trained for natural language tasks. Since the collected dataset consists of solely of Java code, the candidate models were further constrained to  those which were pre-trained on a dataset including Java code. This allows the model to leverage  its prior knowledge of Java code structure.

\noindent \textbf{Performance on Similar Tasks}. To achieve the best possible results, candidate models were limited to those that have performed well in similar tasks, such as defect detection and vulnerability detection. For instance, CodeBERT~\cite{feng2020codebert} achieves good results as part of LineVul~\cite{fu2022linevul}, and GPT-4~\cite{achiam2023gpt} outperforms other models in Gao et al.'s study~\cite{gao2023far}.

\noindent \textbf{Model Size}. The size of the model that could be utilized in this study was limited due to the lack of access to large GPUs for fine-tuning. For instance, CodeLlama~\cite{roziere2023code}, a model that was considered for this study, could not be fine-tuned on the available GPUs due to memory constraints. 

Based on the above criteria, CodeBERT~\cite{feng2020codebert} was selected as the model of interest for this study. This model has been successfully used in other studies to detect vulnerabilities~\cite{fu2022linevul}, \cite{zhou2023devil}. Specifically, we use the ``microsoft/codebert-base'' model available on HuggingFace\footnote{https://huggingface.co/microsoft/codebert-base}.

\subsection{Research Question}
In order to evaluate the approach, we aim to address the following research questions (RQs).

\begin{itemize}[topsep=0pt]
    \item \textbf{RQ1} How does the selected Code Language Model (CLM) perform in detecting code issues (Overall performance)? The primary objective of this study is to evaluate the capability of CLMs. Thus, in this RQ, initially, we examine whether the CLM can detect the presence or absence of issues (binary classification). Subsequently, we assess its ability to identify the types of issues as well (multi-label classification).
    
    \item \textbf{RQ2} How do the different input formulations affect the model's performance (Impact of different input formulation)? As discussed in Section~\ref{inputSection}, there are certain types of information (such as comments) in the target methods that could either be helpful or detrimental to the model. This research question aims to determine how the inclusion or exclusion of different types of information affects the model's performance.
    
    \item \textbf{RQ3} How does the fine-tuned model perform when identifying issues of types that are rare in the dataset (Detection of rare issues)? Due to the nature of source code issues, the collected dataset has a very unbalanced label distribution, with some labels appearing as little as once. This research question investigates whether the model  can reliably identify the presence of these rare issues, which would greatly increase its usefulness in the real world.
    \begin{itemize}
      \item \textbf{RQ3.1} How does the performance of the multi-label classification model differ on rare issue types compared to common issue types? The key goal of the multi-label classification model is to identify the presence of particular issue types. If the model could only identify the commonly found issue types, it would be of limited use.
      \item \textbf{RQ3.2} How does the performance of each model change as rare issues are removed from the dataset? This research question aims to explore the extent to which rarely occurring issues act as confusing noise to the model. 
    \end{itemize}
    
    \item \textbf{RQ4} How does our model's performance differ when analyzing projects that are present in CodeSearchNet compared to those that are not (generality of the model)? Given that CodeBert was pre-trained using the CodeSearchNet dataset~\cite{feng2020codebert}, we want to determine how the pre-training affects the overall performance and whether the presence of projects in both the pre-training and fine-tuning impacts the  model's overall performance. 
    \begin{itemize}
      \item \textbf{RQ4.1} How does the model perform when evaluated on projects that were not included in the fine-tuning dataset? Knowing how well the model generalizes to unseen projects will inform how easily this approach could be adopted in the real-world. 
    \end{itemize}
\end{itemize}

\subsection{Evaluation Metrics}
In order to compare the performance of issue identification, we use the accuracy, precision, recall, and F1 score metrics, which are  commonly used in similar studies~\cite{gao2023far}, \cite{zhou2023devil},  \cite{li2021vuldeelocator}. In the following descriptions, \textit{T} refers to the test dataset, \textit{TP} refers to the number of true positive results, \textit{TN} refers to the number of true negative results, \textit{FP} refers to the number of false positive results, and \textit{FN} refers to the number of false negative results.

\noindent \textbf{Accuracy}: The proportion of test samples for which the model gave the correct output. 
\begin{equation*}
\text{Accuracy} = \frac{TP + TN}{|T|}
\end{equation*}
\noindent \textbf{Precision}: The proportion of true positive predictions amongst all the positive predictions that the model generated.
\begin{equation*}
\text{Precision} = \frac{TP}{TP + FP}
\end{equation*}
\noindent \textbf{Recall}: The proportion of true positive predictions amongst all the positive samples in the test set. 
\begin{equation*}
\text{Recall} = \frac{TP}{TP + FN}
\end{equation*}
\noindent \textbf{F1 Score}: The harmonic mean of the precision and recall. 
\begin{equation*}
\text{F1} = 2 \frac{Precision \cdot Recall}{Precision + Recall}
\end{equation*} 

\subsubsection{Multi-Label Classification Metrics:}
For multi-label classification, the model's precision, recall, and F1 score are reported as the weighted average of each metric across all issue types. For instance, the reported precision value for the multi-label classification model is the weighted average of the model's precision for each issue type. This is calculated using the following formula, where \textit{I} is the set of issue types that appear at least once in the test dataset, \textit{n(i)} is the number of occurrences of issue type \textit{i} in the test dataset, and \textit{m(i)} is value for the metric of interest for the given issue type \textit{i}. 
\begin{equation*}
\text{Weighted Average Metric} = \frac{\sum_{i \in I}n(i)m(i)}{\sum_{i \in I}n(i)}
\end{equation*}
The multi-label classification model's accuracy is calculated as the proportion of test samples where the set of predicted issue types exactly match the set of expected issue types.

\subsection{Overall Models' Performance  (RQ1)}
To answer RQ1, we fine-tune and evaluate a binary classification model and a multi-label classification model for issue detection and issue identification, respectively. For this initial experiment, we use the unmodified target methods, without any of the modifications described in Section~\ref{inputSection}. The dataset for this experiment is split as described in Section~\ref{sec:defaultSplits}, in that the train, validation, and test datasets consist of 80\%, 10\%, and 10\% of the total dataset, respectively. The results of this experiment can be seen in Table~\ref{table:lintingEffectiveness}.

As can be seen, the approach achieves a high degree of accuracy in both tasks. The binary classification model achieves an accuracy of 0.840  with an F1 score of 0.842 when detecting issues. Similarly, the multi-label classification can identify issues with an accuracy of 0.832 and an F1 score of 0.838. 

When comparing these results to other works, we outperform the results reported by Zhou et al. (an accuracy of 0.731)~\cite{zhou2023devil}, and Yuan et al. (an accuracy of 0.61)~\cite{yuan2023evaluating}. While the work by Fu and Tantithamthavorn outperforms ours (as they achieved an F1 score of 0.91), their dataset focuses solely on vulnerabilities, whereas ours is broader, including many other classes of issues such as poor performance (e.g., issues \textit{I4} and \textit{S66}), poor code organization (e.g., issue \textit{S21}), and poor code clarity (e.g., issue \textit{S16}), in addition to issues related to vulnerabilities (e.g., issues \textit{S23} and \textit{S40}).

\begin{table}[!htb]
    \centering
    \begin{tabular}{l c c c c}
        \toprule
        Model & Accuracy & Precision & Recall & F1 \\
        \midrule
        Binary & 0.840 & 0.828 & 0.857 & 0.842 \\
        Multi-Label & 0.832 & 0.851 & 0.834 & 0.838 \\
        \bottomrule
    \end{tabular}
    \caption{The Effectiveness of the Binary Classification and Multi-Label Classification Models at Linting}
    \label{table:lintingEffectiveness}
\end{table}

\subsubsection{Comparison to Linting Tools}
We compare the model's performance with the linters used for data collection (Infer and SpotBugs). For this comparison, we consider all the methods in the test dataset that have at least one issue. Of these methods, 25.0\% are flagged as having an issue by Infer, 82.7\% are flagged by SpotBugs, and our approach correctly identifies 85.7\% of the methods with issues (as seen through the recall reported in Table~\ref{table:lintingEffectiveness}). This shows that our approach is capable of identifying more methods with issues than either tool can by itself.

Additionally, we compare the amount of time it takes to run each of the linters  compared to our approach. Due to the required length of time to perform the analysis, it was infeasible to re-analyze all 6,477 candidate projects with both linters. Instead, each linter was run on a number of randomly selected projects which comprised 10\% of the data collected by each linter. This means that 103 projects were analyzed for Infer and 182 projects were analyzed for SpotBugs. The fine-tuned models were timed on the according to how long it took to format and  analyze each method in the selected projects. To ensure a fair comparison, each model was evaluated twice: once on the projects analyzed by Infer, and once on the projects analyzed by SpotBugs. In addition, the time it took to extract every method in the selected projects was recorded, which is our approach's equivalent to the linters' requirement of compiling the code.

\begin{figure}[!t]
    \centering
    \includegraphics[width=\columnwidth]{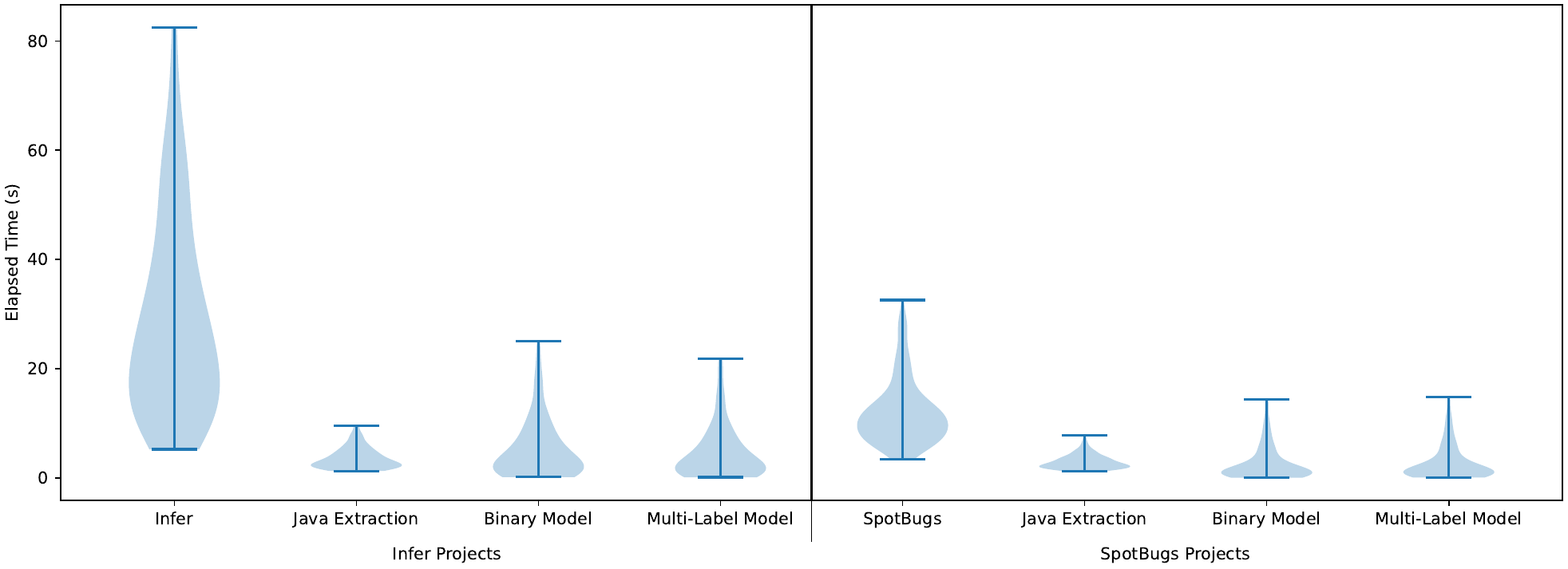}
    \caption{Plot Showing the Times that Each Part of Our Approach Takes for Each Component on the Infer and SpotBugs Projects (on the Left and Right of the Plot Respectively). Outliers are Excluded from this Plot.}
    \label{fig:approachTimes}
\end{figure}

A summary of the results of this evaluation can be seen in Figure~\ref{fig:approachTimes}. On the 103 projects selected for Infer, Infer took an average of 51.395 seconds to analyze a project. In comparison, the extraction of the Java source code took an average and 4.583 seconds to extract all methods, the binary classification model took an average of 11.310 seconds to analyze all methods, and the multi-label classification model took an average of 9.325 seconds to analyze all methods. Our approach is faster than  Infer, requiring 51.11\% less time than Infer's analysis to run the Java extraction, binary model, and multi-label model in sequence. 

For the 182 Spotbugs projects, SpotBugs took an average of 27.203 seconds to analyze each project. Extraction of the Java source code took an average and 4.376 seconds to extract all methods, the binary model took an average of 7.817 seconds to analyze all methods, and the multi-label model took an average of 8.429 seconds to analyze all methods. This shows that our approach is much faster than applying SpotBugs, requiring 24.20\% less time than SpotBugs' analysis to run the Java extraction, binary model, and multi-label model in sequence.

The most significant performance improvement comes when comparing our approach to running both tools, in which case our approach is 67.92\% faster. This figure was determined by summing the average times for both linters (51.395 seconds for Infer and 27.203 seconds for SpotBugs), and comparing that to the sum of the average times for our approach on the Infer projects (25.218 seconds total). We only use the times for our approach on the Infer projects for this comparison, since they are the slower set of times. This shows that, in addition to  finding a higher ratio of the issues in the code than either individual linter, it is also much faster at analyzing a codebase than the linters. 

It is worth noting that, in addition to prediction time, the model also requires tuning, which is not considered in the above comparison. This is because training does not occur continuously and can be done offline, whereas the comparison focuses on the analysis required during CI cycles.

\begin{conclusionBox}{RQ1 Conclusion}
Our approach was able to achieve performances which are comparable to the performances achieved by state of the art approaches in a similar field of study, vulnerability detection. In addition, our approach detects more issues than either code linter can by itself while taking less time to analyze a project on average. 
\end{conclusionBox}

\subsection{Input Format Comparison (RQ2)}  \label{rq2Eval}
This section aims to evaluate how the models perform with  different input formats. These input formats are created by applying the operations described in Section~\ref{inputSection} in different combinations. The dataset for these experiments is split as described in Section~\ref{sec:defaultSplits}, and the same splits were used for each experiment. The results of this evaluation can be seen in Table~\ref{table:bcResults} for binary classification and Table~\ref{table:mcResults} for multi-label classification. In these tables, a shorthand is used to denote the applied input formulation operations. \textit{RC} represents the remove comments operation, \textit{RJ} represents the remove Javadoc operation, and \textit{RS} represents the replace string literals operation. The \textit{unmodified} format is the target method input with no operations applied. The results of this experiment can be seen in Table~\ref{table:bcResults} and Table~\ref{table:mcResults}, for the binary classification model, and the multi-label classification model respectively. The best performing formats are the \textit{RJ} format for the binary classification problem, achieving an accuracy of 0.849 and an F1 score of 0.847. The \textit{RC+RJ} format performed the best for multi-label classification problem, with an accuracy of 0.836 and F1 score of 0.843.

Interestingly, in both cases the best performing input formats are still fairly close to the performance of the unmodified input. For binary classification, the unmodified input achieves an accuracy of 0.840 and an F1 score of 0.842, only slightly lagging behind the \textit{RJ} format, and achieves the best recall score, being 0.857. This likely indicates that while some benefit can be gained from the operations  applied to the target method, additional context (such as the methods which the target method calls) is likely required to gain significant performance improvements beyond what is achieved in this study. Exploring this idea is left for future work.

In the case of multi-label classification, all of the results are close together, with the accuracy metrics having only a difference of 1.2 percentage points between the best and worst accuracy measures (the \textit{RC+RJ} and \textit{RS} inputs respectively), with the unmodified format achieving an accuracy of 0.832 (0.004 less than the highest accuracy). This is likely due to the unbalanced nature of the dataset's issue types, where the lesser occurring issue types may be limiting the model's performance. This idea is explored more in Section~\ref{rq2Eval}. 

The F1 scores for each issue type were also investigated to determine if particular issue types had better performance with certain input formats compared to the average. Overall, the variance in F1 score results between different input formats seems to be partially tied to the number of instances of the issue type in the dataset. For each of the 10 most frequently occurring issue types in the test dataset, the best and worst F1 scores have a difference of less than 0.061. In contrast, the less numerous issue types can have a difference of up to 0.821 between the best and worst F1 scores,  indicating that the input format is less important than the number of samples for an issue type. Overall, 6 issue types (including 4 of the top 5 most numerous types) achieve their best F1 performance using the \textit{RC+RJ} format, matching the best format according to the weighted average. 

Further analysis was performed on all issue types that had a difference greater than 0.01 between their best and worst F1 scores (as a difference less than this is interpreted as the model's performance on that issue type being input-format agnostic). Interestingly, every input format resulted in the best F1 score for at least one issue type. The performance of different input formats does not seem to follow any trends within issue types. For instance, consider issue \textit{S18}, for which the \textit{RC+RJ} format achieved the best F1 score of 0.627, while the \textit{RC} and \textit{RJ} formats resulted in some of the worst F1 scores (0.368 and 0.412 respectively) for this type. Based on a manual analysis of the issue types and some samples from the dataset, it was determined that none of the input formats would be particularly helpful over the others when manually identifying issues. Many issue types (such as \textit{S15} and \textit{S23}) would require additional context from either the target method's class or from the methods which the target method calls. Based on this analysis, an avenue for future work would be to investigate how our approach performs when using input formats that include additional context from outside the target method. 

\begin{conclusionBox}{RQ2 Conclusion} 
For the binary classification model, it was determined that the \textit{RJ} input format performed the best. It achieved the best results in both the accuracy and F1 metrics, and nearly matches the best precision result. It does lag behind other formats in recall. For the multi-label classification, the \textit{RC+RJ} format achieves the best performance, achieving or matching the best result in every metric except precision, where it very slightly lags behind. Given these results, the remaining experiments are performed using these input formats for each of the respective models. 
\end{conclusionBox}

\begin{table}[!htb]
    \centering
    \begin{tabular}{l c c c c}
        \toprule
        Input Format & Accuracy & Precision & Recall & F1 \\
        \midrule
        Unmodified & 0.840 & 0.828 & \textbf{0.857} & 0.842 \\
        RC & 0.818 & 0.795 & \textbf{0.857} & 0.825 \\
        RJ & \textbf{0.849} & 0.854 & 0.841 & \textbf{0.847} \\
        RS & 0.826 & 0.845 & 0.798 & 0.821 \\
        RC+RJ & 0.842 & 0.835 & 0.852 & 0.844 \\
        RC+RS & 0.828 & 0.814 & 0.850 & 0.832 \\
        RJ+RS & 0.836 & \textbf{0.855} & 0.809 & 0.831 \\
        RC+RJ+RS & 0.819 & 0.807 & 0.838 & 0.822 \\
        \bottomrule
    \end{tabular}
    \caption{Binary Classification Results}
    \label{table:bcResults}
\end{table}

\begin{table}[!htb]
    \centering
    \begin{tabular}{l c c c c}
        \toprule
        Input Format & Accuracy & Precision & Recall & F1 \\
        \midrule
        Unmodified & 0.832 & 0.851 & 0.834 & 0.838 \\
        RC & 0.832 & \textbf{0.854} & 0.833 & 0.838 \\
        RJ & 0.830 & 0.852 & 0.833 & 0.837 \\
        RS & 0.824 & 0.845 & 0.826 & 0.831 \\
        RC+RJ & \textbf{0.836} & 0.853 & \textbf{0.838} & \textbf{0.843} \\
        RC+RS & 0.827 & 0.851 & 0.829 & 0.835 \\
        RJ+RS & 0.827 & \textbf{0.854} & 0.833 & 0.839 \\
        RC+RJ+RS & 0.827 & 0.851 & 0.829 & 0.835 \\
        \bottomrule
    \end{tabular}
    \caption{Multi-label Classification Results}
    \label{table:mcResults}
\end{table}

\subsection{Classification Performance on Rare Issue Types (RQ3)} \label{rq3Eval}
This section address \textbf{RQ3}, which examines  how the models perform when confronted with rare issue types. Two experiments were performed to address this research question. Note that analyzing rare issues (RQ2.1) is only conducted for multi-class classification since our dataset includes sufficient samples of both classes for binary classifications.

\subsubsection{Analyzing the Multi-Label Classification Model's Performance on Rare Issue Types (RQ3.2):}  
To analyze how well the multi-label classification model performs on rare issue types, we adopt Zhou et al.'s approach which considers so called ``head'' and ``tail'' labels separately~\cite{zhou2023devil}. This approach is designed to analyze the effectiveness of language models on datasets where the labels follow a long-tail distribution (i.e., a dataset where relatively few of the labels make up the majority of the samples). This is done by analyzing the ``head'' labels and ``tail'' labels separately, with ``head'' labels being the  most frequently occurring labels that account for 50\% of the dataset, and the ``tail'' labels being the less frequent labels that account for the remaining 50\% of the dataset. 

Note that, we  deviate from Zhou et al.'s methodology in how we handle label overlap. Label overlap occurs when a label could be classified as either a head label or tail label (such as a case where 45\% of the samples are head samples, 40\% are tail samples, and the remaining 15\%  all have the same label). Zhou et al. allocate the overlapping label to minimize the difference between the number of head samples and tail samples~\cite{zhou2023devil}. In our dataset, 50\% of the samples have no issues, and are labeled as such. If we follow  Zhou et al.'s methodology exactly, the only head label would be the ``no issues'' label. However, the next most frequently occurring label is for issue \textit{S8}, with 25.4\% of samples having this label. To provide a more accurate view of how our model performs on the infrequently occurring labels, we include the \textit{S8} label in the head labels (bringing our head label percentage to 75\% instead of 50\%). 

This analysis was performed using \textit{RC+RJ} results from Table~\ref{table:mcResults}. The model's performance on these two groups of labels can be seen in Table~\ref{table:mcHeadVsTail}. While the two groups have a comparable accuracy (0.887 for head labels and 0.875 for tail labels), the head label performance on each of the other metrics far exceeds the tail label performance. The tail label accuracy is skewed by the fact that 75.3\% of the samples have no expected tail labels. When  considering only samples that have at least one expected tail label, the model's accuracy on tail labels becomes 0.598. This low performance on the tail labels show that our model suffers from the unbalanced dataset, and fails to learn the patterns that indicate rarer issues.

To analyze how our approach performs when utilizing a more balanced (although still unbalanced) dataset, this analysis was repeated using a pipeline composed of our binary classification model, followed by our multi-label classification model. For this experiment, the dataset was split into training, validation and test datasets using an 80:10:10 split, as described in Section~\ref{sec:defaultSplits}. The binary classification model was fine-tuned using the entirety of the training and validation datasets, while the multi-label classification model was  fine-tuned only using the methods with issues contained in those datasets. The target methods were formulated using the \textit{RJ} format for the binary classification model, and the \textit{RC+RJ} format for the multi-label classification model, following  the results reported in Section~\ref{rq2Eval}. The pipeline was then tested using the algorithm defined in Algorithm~\ref{fig:pipelineAlg}. As can be seen, the multi-label classification model is only used to evaluate the target methods that the binary classification model identifies as having an issue. 

\begin{algorithm}
\small
\caption{Evaluating a Pipeline Which Utilizes both the Binary Model and the Multi-Label Model}
 \label{fig:pipelineAlg}
\begin{algorithmic}[1]
    \State binaryResults $\gets$ []
    \State multiLabelResults $\gets$ []
    \For{targetMethod $in$ testDataset}
        \State formattedMethod $\gets$ formatMethod(targetMethod,$ 'RJ'$)
        \State binResult $\gets$ binaryModel.evaluate(formattedMethod)
        \State binaryResults.append(binResult)
        \If{indicatesIssue(binResult)}
            \State formattedMethod $\gets$ formatMethod(targetMethod,$ 'RC+RJ'$)
            \State multiResult $\gets$ multiLabelModel.evaluate(formattedMethod)
            \State multiLabelResults.append(multiResult)
        \EndIf
    \EndFor
    \end{algorithmic}   
\end{algorithm}

\begin{table}[!htb]
    \centering
    \begin{tabular}{l c c c c}
        \toprule
        Label Type & Accuracy & Precision & Recall & F1 \\
        \midrule
        \multicolumn{5}{l}{\textbf{All Methods}} \\
        \midrule
        All Data & 0.836 & 0.853 & 0.838 & 0.843 \\
        Head Labels & 0.887 & 0.905 & 0.919 & 0.912 \\
        Tail Labels & 0.875 & 0.703 & 0.607 & 0.644 \\
        \midrule
        \multicolumn{5}{l}{\textbf{Pipeline}} \\
        \midrule
        Binary Model & 0.872 & 0.872 & 0.872 & 0.872 \\
        \midrule
        All Labels & 0.764 & 0.789 & 0.847 & 0.811 \\
        Head Labels & 0.924 & 0.896 & 0.985 & 0.938 \\
        Tail Labels & 0.827 & 0.589 & 0.591 & 0.574 \\
        \bottomrule
    \end{tabular}
    \caption{Comparison of Multi-label Classification Model Performance on Head Labels Compared to Tail Labels}
    \label{table:mcHeadVsTail}
\end{table}

This approach was used instead of simply fine-tuning and testing the multi-label classification model only on methods with issues  to ensure a fair comparison to the results obtained when fine-tuning and testing the model on all methods. The results of this experiment are shown in Table~\ref{table:mcHeadVsTail} (which also shows the performance of the pipeline's binary classification model). As can be seen, the multi-label classification model in the pipeline overall achieves a lower performance when considering all labels, with an accuracy of 0.764 and F1 score of 0.811, compared to the baseline experiment's accuracy of 0.836 and F1 score of 0.843. Comparatively, the pipeline's model performs better on the head labels, and worse on the tail labels.

\subsubsection{Model Performance With Different Minimum Issue Count Thresholds (RQ3.2):}
This research question examines  how the model's performance changes as the dataset is filtered to remove less frequently occurring issues. For this comparison, four datasets were utilized, each of which filters out issues based on a different frequency of occurrence in the overall dataset. The first dataset includes all issues retrieved during data collection without any filtering. The second dataset is the baseline dataset used throughout this paper, containing the top 75\% most frequently occurring issues (issues that occur at least 15 times). The third dataset includes only the top 50\% of issues,  occurring at least 47 times. Finally, the last dataset contains the top 20\% of issues, each  occurring at least 601 times. In each case, methods without issues were randomly selected for inclusion in the datasets such that there is an equal number of methods with issues and methods without issues. The details of each dataset can be found in Table~\ref{table:thresholdDatasets}.

\begin{table}[!htb]
    \centering
    \begin{tabular}{p{0.1\linewidth} p{0.1\linewidth} p{0.1\linewidth} p{0.1\linewidth} p{0.1\linewidth} p{0.11\linewidth} p{0.15\linewidth}}
        \toprule
        Included Types & Number of Types & Methods With Issues & Train Set Size & Test Set Size & Binary Accuracy & Multi-Label Accuracy \\
        \midrule
        All Types & 100 & 84747 & 135598 & 16948 & 0.843 & 0.824 \\
        Top 75\% & 75 & 84664 & 135464 & 16932 & 0.849 & 0.836 \\
        Top 50\% & 50 & 84163 & 134662 & 16832 & 0.848 & 0.834 \\
        Top 20\% & 20 & 80652 & 129044 & 16130 & \textbf{0.851} & \textbf{0.866} \\
        \bottomrule
    \end{tabular}
    \caption{Dataset Stats for Different Thresholds}
    \label{table:thresholdDatasets}
\end{table}

The results of this experiment can be seen in Table~\ref{table:binaryThresholdResults} and Table~\ref{table:multiThresholdsResults}. As  expected, the performance of both models tends to increase as the less frequently occurring issue types are filtered out. The binary classification model achieves an accuracy of 0.843 when considering all issues, and 0.851 when considering the top 20\% of issue types. The multi-label classification model achieves an accuracy of 0.824 when considering all issues, and 0.866 when considering the top 20\% of issue types. However, the binary model's precision peaks with dataset containing the top 75\% of issue types. Interestingly, in both cases the dataset for the top 50\% of issue types results in a model performance that is very similar to (or even slightly worse than) the results from the dataset for the top 75\% of issue types.

\begin{table}[!htb]
    \centering
    \begin{tabular}{l c c c c}
        \toprule
        Included Issues & Accuracy & Precision & Recall & F1 \\
        \midrule
        All Issue Types & 0.843 & 0.839 & 0.850 & 0.844 \\
        Top 75\% & 0.849 & \textbf{0.854} & 0.841 & 0.847  \\
        Top 50\% & 0.848 & 0.837 & 0.864 & 0.851 \\
        Top 20\% & \textbf{0.851} & 0.832 & \textbf{0.879} & \textbf{0.855} \\
        \bottomrule
    \end{tabular}
    \caption{Binary Classification Results for Different Issue Type Cutoffs}
    \label{table:binaryThresholdResults}
\end{table}

\begin{table}[!htb]
    \centering
    \begin{tabular}{l c c c c}
        \toprule
        Included Issues & Accuracy & Precision & Recall & F1 \\
        \midrule
        All Issue Types & 0.824 & 0.842 & 0.828 & 0.830 \\
        Top 75\% & 0.836 & 0.853 & 0.838 & 0.843 \\
        Top 50\% & 0.834 & 0.855 & 0.838 & 0.842 \\
        Top 20\% & \textbf{0.866} & \textbf{0.884} & \textbf{0.871} & \textbf{0.875} \\
        \bottomrule
    \end{tabular}
    \caption{Multi-label Classification Results for Different Issue Type Cutoffs}
    \label{table:multiThresholdsResults}
\end{table}

\begin{conclusionBox}{RQ3 Conclusion} 
The work presented in this section draws a clear conclusion that our models are much more effective at identifying more common issue types. This falls within the results of previous work performed by Zhout et al.~\cite{zhou2023devil} which showed that NLMs are much more effect at identifying common labels in a dataset. In particular this shows the importance of collecting a well-balanced dataset when approaching the task of code linting with NLMs.
\end{conclusionBox}

\subsection{Generality of the model (RQ4)}

As specified by Feng et al., CodeBERT was pre-trained on the CodeSearchNet (CSN) dataset~\cite{feng2020codebert}. Since the majority of our dataset was created using Java projects  included in the CodeSearchNet dataset, the models' results were analyzed to determine if there was a difference in performance when considering projects that were in the model's pre-training dataset and those that are not.

In the test dataset, there are 16,932 total methods. 14,693 of those methods are from projects included in the CodeSearchNet dataset, while the remaining 2,239 are not. The differences in performance can be seen in Table~\ref{table:bcDataSource} for the binary classification model, and Table~\ref{table:mcDataSource} for the multi-label classification model. In both cases, the model performed better across all metrics on the projects which are included in the CodeSearchNet dataset, although the difference in performance was  more significant in the multi-label classification model than the binary classification model. This shows that in both cases, pre-training information retained by the model about the projects  helps in identifying issues. 

\begin{table}[!htb]
    \centering
    \begin{tabular}{l c c c c}
        \toprule
        Test Set Contents & Accuracy & Precision & Recall & F1 \\
        \midrule
        Full Test Set & 0.849 & 0.854 & 0.841 & 0.847 \\
        CSN Projects & \textbf{0.852} & \textbf{0.857} & \textbf{0.845} & \textbf{0.851} \\
        Non-CSN Projects & 0.824 & 0.832 & 0.820 & 0.826 \\
        \bottomrule
    \end{tabular}
    \caption{Comparison of Binary Classification Results Using Test Sets Consisting of Projects Retrieved from CodeSearchNet and from the GitHub API}
    \label{table:bcDataSource}
\end{table}

\begin{table}[!htb]
    \centering
    \begin{tabular}{l c c c c c}
        \toprule
        Test Set Contents & Accuracy & Precision & Recall & F1 \\
        \midrule
        Full Test Set & 0.836 & 0.853 & 0.838 & 0.843 \\
        CSN Projects & \textbf{0.844} & \textbf{0.859} & \textbf{0.846} & \textbf{0.849} \\
        Non-CSN Projects & 0.785 & 0.812 & 0.791 & 0.798 \\
        \bottomrule
        \end{tabular}
    \caption{Comparison of Multi-Label Classification Results Using Test Sets Consisting of Projects Retrieved from CodeSearchNet and from the GitHub API}
    \label{table:mcDataSource}
\end{table}

\subsubsection{Model Performance on Unseen Projects (RQ4.1) during fine-tuning:}
To determine how well the models generalize to unseen projects during fine-tuning, the models were tested on projects that were withheld from the fine-tuning dataset. Projects to be withheld were randomly selected until 10\% of the dataset was selected. This random selection was done in such a way that half of the selected projects were in CodeSearchNet, and the other half were not. This created a test set containing 16,952 methods. The remainder of the dataset was then split into a training, validation, and test datasets (containing 80\%, 10\%, and 10\% of the  remaining dataset, respectively). This second test dataset (containing projects that the models are fine-tuned on) was used as a baseline for comparison. The results of this experiment can be seen in Table~\ref{table:bcUnseenProjects} and Table~\ref{table:mcUnseenProjects} for the binary classification model and multi-label classification model, respectively. Given this setup, there are four cases to be analyzed:
\begin{enumerate}
    \item Projects present in both the pre-training and fine-tuning datasets (\textit{Seen CSN Projects})
    \item Projects present only in the fine-tuning dataset (\textit{Seen Non-CSN Projects})
    \item Projects present only in the pre-training dataset (\textit{Unseen CSN Projects})
    \item Projects that are present in neither the pre-training nor fine-tuning datasets (\textit{Unseen Non-CSN Projects})
\end{enumerate}

First, we compare cases 1 and 3 to cases 2 and 4 respectively, to determine the effect of projects being included in the pre-training dataset. As expected, the performance on the projects used in pre-training surpasses performance on the projects not used for pre-training, regardless of presence in the fine-tuning dataset (which is consistent with the results presented in Table~\ref{table:bcDataSource} and Table~\ref{table:mcDataSource}). 

Next, we can compare case 1 to case 3, and case 2 to case 4 to analyze the effect of projects being present in the fine-tuning dataset. Considering cases 1 and 3, we find some particularly interesting results, in that presence in the fine-tuning dataset does not seem to offer any particular advantage. In the case of the binary classification model, case 3 outperforms case 1 in every metric, achieving an accuracy of 0.866 and F1 score of 0.865, compared to case 1's accuracy of 0.831 and F1 score of 0.827. For the multi-label classification model, cases 1 and 3 give very similar results, with neither having an overall better performance, case 1 having an accuracy of 0.828 and F1 score of 0.836 while case 3 has an accuracy of 0.838 and F1 score of 0.832. Similarly, cases 2 and 4 also yield similar results, especially for the binary classification model. For this model, case 2 achieves an accuracy of 0.798 and F1 score of 0.790 while case 4 achieves an accuracy of 0.7991 and F1 score of 0.775. The difference is more pronounced in the multi-label classification model, where case 2 outperforms case 4 by 0.028 in terms of accuracy and 0.043 in terms of F1 score. Overall, this analysis shows that presence in the fine-tuning dataset does not necessarily result in improved performance.

When considering projects not in the pre-training dataset (cases 2 and 4), there is a decrease in performance in both models from cases 1 and 3. This  likely  indicates that the model struggles more with projects that it has never been exposed to, which aligns with expectations. However, this decrease is not  substantial, and the models still perform well on the unseen projects. For instance, the binary model achieves an accuracy of 0.798 and an F1 score of 0.790 on case 2, and an accuracy of 0.791 and an F1 score of 0.775 on the case 4. The multi-label model shows a larger performance difference, achieving an accuracy of 0.750 and an F1 score of 0.785 on case 2 and an accuracy of 0.722 and an F1 score of 0.742 on the case 4. Overall, these results  indicate that our approach generalizes well to unseen projects, although it does perform better on projects that were used during pre-training and fine-tuning.

\begin{table}[!htb]
    \centering
    \begin{tabular}{c l c c c c}
        \toprule
        Case & Test Set Contents & Accuracy & Precision & Recall & F1 \\
        \midrule
        1 & Seen CSN Projects & 0.831 & 0.846 & 0.810 & 0.827 \\
        2 & Seen Non-CSN Projects & 0.798 & 0.830 & 0.754 & 0.790 \\
        3 & Unseen CSN Projects & \textbf{0.866} & \textbf{0.891} & \textbf{0.841} & \textbf{0.865} \\
        4 & Unseen Non-CSN Projects & 0.791 & 0.823 & 0.733 & 0.775 \\
        \bottomrule
    \end{tabular}
    \caption{Comparison of Binary Classification Performance on Test Sets Containing Either Seen or Unseen Projects}
    \label{table:bcUnseenProjects}
\end{table}

\begin{table}[!htb]
    \centering
    \begin{tabular}{c l c c c c}
        \toprule
        Case & Test Set Projects & Accuracy & Precision & Recall & F1 \\
        \midrule
        1 & Seen CSN Projects & 0.828 & \textbf{0.848} & 0.832 & \textbf{0.836} \\
        2 & Seen Non-CSN Projects & 0.750 & 0.761 & 0.756 & 0.785 \\
        3 & Unseen CSN Projects & \textbf{0.838} & 0.833 & \textbf{0.840} & 0.832 \\
        4 & Unseen Non-CSN Projects & 0.722 & 0.732 & 0.724 & 0.742 \\
        \bottomrule
    \end{tabular}
    \caption{Comparison of Multi-Label Classification Performance on Test Sets Containing Either Seen or Unseen Projects}
    \label{table:mcUnseenProjects}
\end{table}

\begin{conclusionBox}{RQ4 Conclusion} 
While the model performs better on the projects of interest when they are included in the pre-training dataset, it is much less impactful to exclude a project from the fine-tuning dataset. This shows that our approach generalizes well to projects it was not fine-tuned on, as it performs nearly as well on projects that were included in the fine-tuning dataset as those that were not. Closing the gap between projects that were present in the pre-training dataset compared to those that were not could perhaps be done by using a larger NLM or a different training methodology, however this is left for future work.
\end{conclusionBox}
\section{Threats to Validity}
\noindent \textbf{Potential False Positives and False Negatives in Dataset}: During data collection, an implicit assumption was made that the results generated by the linting tools were correct (i.e. that the reported issues are actually issues and are all of the issues in the project). However, it is very likely that this is not the case. Previous work has found that Infer has a 72.7\% precision for issue \textit{I1} and a 57.4\% precision for issue \textit{I3}~\cite{kharkar2022learning}. Unfortunately, precision figures for other Infer issue types and the SpotBugs tool were not readily available, and there seems to be no reported results about potential false negatives in the tools' output (i.e. an issue of a type that the tool supports and fails to report). Presence of false positives and false negatives in the dataset would prevent the models from effectively learning the patterns that comprise true issues, and would instead cause it to learn to replicate the tools' performance. However, the primary focus of this study is to examine the model's capability in identifying issues given a dataset of code snippets and labels. Applying these techniques in production may require a more precise examination of the training dataset to ensure its high quality.

\noindent \textbf{Dataset Imbalance}: As previously discussed in Section~\ref{rq3Eval}, the dataset utilized in this study is very imbalanced, being dominated by relatively few issue types. While the dataset is balanced so that half of the code snippets have no issues, even if those are not considered, more than half of the identified issues are of two issue types (issues \textit{S8} and \textit{I5}). The imbalanced dataset likely led to the models learning to identify the the common issue types while treating the very rarely occurring ones as noise. A more balanced dataset would help reduce any bias that the models have towards certain issue types, and may lead to better performance on rarely occurring issue types. In order to determine the impact of this imbalance on our approach, we analyzed the performance of our model on both the common and rare issue types separately, as well as analyzed how the performance of our model changes as the dataset became less imbalanced. We found that our model does perform better on the more common issue types, and that its overall performance improves as the dataset becomes more balanced. In future work, a potential way to alleviate this would be to augment a dataset built from natural code with synthetically created issues.
\vspace{-.2cm}
\section{Conclusion}
This paper explores the potential of leveraging large language models  for code linting with the aim of  addressing several limitations inherent in traditional linters. The investigation into the architecture and implementation of a language model-based code linter suggests that large language models can offer improvements in accuracy and performance. The empirical evaluation through the research questions provides preliminary evidence of the effectiveness of large language models in code linting. The results suggest that our language model-based approach can achieve competitive accuracy and can outperform existing linters in speed. The analysis of different input formulations underscores the importance of optimizing input data to enhance model performance.

However, our study also identifies areas where large language models face challenges, such as detecting rare issues in the dataset and analyzing projects not included in the pre-training data. These findings highlight the need for more diverse and comprehensive training datasets and further research to improve the model's ability to generalize across different types of codebases and issue types. Overall, while this work contributes to the understanding of large language models' application in software engineering, it is an initial step in exploring their potential for creating more efficient, accurate, and adaptable code linting tools. As large language models continue to evolve, we hope that further advancements will lead to more robust and maintainable software systems.

%

%



%
%

\bibliographystyle{spmpsci}      
\bibliography{refrences}   

\newpage

\begin{appendices}
\section{Collected Issue Types}

\begin{longtable}{p{0.05\linewidth} | p{0.25\linewidth} | p{0.005\linewidth} | p{0.35\linewidth}}
    \caption{The list of issue types retrieved during dataset collection, and a brief description of each} \label{tab:long} \\ \hline \hline
    
    \hline \textbf{ID} & \textbf{Issue} & \textbf{Count} & \textbf{Description} \\ \hline \hline \hline 
    \endfirsthead
    
    \multicolumn{3}{c}%
    {{\bfseries \tablename\ \thetable{} -- continued from previous page}} \\ \hline \hline
    \hline \textbf{ID} & \textbf{Issue} & \textbf{Count} & \textbf{Description} \\ \hline \hline \hline 
    \endhead
    
    \hline \multicolumn{4}{r}{{Continued on next page}} \\ \hline \hline
    \endfoot
    
    \hline \multicolumn{4}{r}{{End of table}} \\ 
    \hline \hline
    \endlastfoot
    
    \multicolumn{4}{l}{\textbf{Infer Issues}~\cite{inferDesc}} \\
    \hline \hline 
    I1 & NULL\_\newline DEREFERENCE & 7135 & Potential null pointer dereference at the specified location \\ \hline
    I2 & NULLPTR\_\newline DEREFERENCE & 121 & Potential null pointer dereference at the specified location \\ \hline
    I3 & RESOURCE\_LEAK & 2133 & A resource may not be closed, especially in the case were an exception occurs while accessing the resource \\ \hline
    I4 & INEFFICIENT\_\newline KEYSET\_\newline ITERATOR & 769 & A HashMap is iterated over using the keySet() iterator instead of the entrySet() iterator \\ \hline
    I5 & THREAD\_\newline SAFETY\_\newline VIOLATION & 9241 & A potential data race between threads \\ \hline
    I6 & EXPENSIVE\_\newline LOOP\_\newline INVARIANT\_ CALL & 26 & An expensive method with constant input and output is repeatedly called in a loop \\ \hline
    I7 & CHECKERS\_\newline IMMUTABLE\_\newline CAST & 319 & A method with a mutable return type returns an immutable object \\ \hline
    I8 & INTERFACE\_NOT\_\newline THREAD\_SAFE & 3246 & An interface not annotated with @ThreadSafe is called from a thread safe context \\ \hline
    I9 & ARBITRARY\_\newline CODE\_\newline EXECUTION\_ UNDER\_LOCK & 2 & While a lock is held, a call to arbitrary code which may obtain a lock is made, potentially resulting in a deadlock \\ \hline
    I10 & DEADLOCK & 76 & Two distinct threads attempt to obtain the same locks in different orders \\ \hline
    \midrule
            \multicolumn{4}{l}{\textbf{SpotBugs Issues}~\cite{spotBugsDesc}} \\
            \hline \hline
    S1 & NP & 1662 & A method may unexpectedly return null, a field or paramater may be null, or a null value is guaranteed to be dereferenced \\ \hline
    S2 & WMI & 748 & A HashMap is iterated over using the keySet() iterator instead of the entrySet() iterator \\ \hline
    S3 & CT & 8120 & A constructor can throw an exception, making it vulnerable to Finalizer attacks \\ \hline
    S4 & DM & 3244 & A method is invoked in an incorrect or questionable way \\ \hline
    S5 & SF & 994 & A switch statement is missing the default case or contains a case which falls through to the next  \\ \hline
    S6 & OBL & 981 & An object requiring explicit clean up is not cleaned up \\ \hline
    S7 & OS & 313 & A stream fails to be closed under certain conditions \\ \hline
    S8 & EI & 47057 & A reference to an object's mutable field is returned, potentially leading to unchecked and unexpected changes to the object's field \\ \hline
    S9 & JLM & 81 & Synchronization on an object is performed incorrectly or in a confusing way \\ \hline
    S10 & REFLC & 94 & Reflection is used in a public method to create a class specified by a parameter, which could increase the visibility of other classes in the package \\ \hline
    S11 & REFLF & 30 & Reflection is used to increase the accessibility of a field \\ \hline
    S12 & REC & 342 & A catch block is used to catch catch Exception objects rather than specific exception types, which may incorrectly catch a thrown RuntimeException \\ \hline
    S13 & UC & 234 & A condition always produces the same value, a method performs no useful work, or an object is created and modified yet yields no side-effects and never leaves the current context \\ \hline
    S14 & RCN & 659 & A null check is performed on a value that is known to be non-null \\ \hline
    S15 & ST & 611 & A static field is modified by an instance method \\ \hline
    S16 & NM & 654 & A class, method, or field name is either confusing, ignores Java naming conventions, or is a keyword from a later Java version \\ \hline
    S17 & RV & 1250 & A return value is ignored or is handled in a way that could yield unexpected behaviour \\ \hline
    S18 & DLS & 1288 & A value is set or modified, and then is never used \\ \hline
    S19 & ML & 25 & Synchronization is performed on an object referenced from a mutable field, potentially leading to threads locking different objects \\ \hline
    S20 & SBSC & 390 & A string is concatenated in a loop using the '+' operator, where a StringBuilder would be more appropriate \\ \hline
    S21 & SIC & 268 & An inner class should be made into a static inner class since it doesn't use its embedded reference to the outer class \\ \hline
    S22 & SSD & 41 & An instance level lock on a static field may not guard against concurrent access \\ \hline
    S23 & MS & 696 & A mutable, final static field could be accidentally or maliciously altered, and should be made immutable or less visible \\ \hline
    S24 & UG & 48 & A synchronized set method has an associated unsynchronized get method, potentially leading to incosnistent caller states \\ \hline
    S25 & DC & 99 & An instance of double-checked locking may be used here, which may function incorrectly or in unexpected ways on some platforms \\ \hline
    S26 & BIT & 36 & A bitwise operator is used in a way that could lead to unexpected behaviour \\ \hline
    S27 & RANGE & 21 & A provided parameter to a string or array access will be out of bounds \\ \hline
    S28 & UL & 43 & A method may not release all obtained locks in some executation paths \\ \hline
    S29 & SE & 327 & A serializable class may be configured in a way that leads to incorrect serialization or deserialization \\ \hline
    S30 & NS & 33 & Non-short-circuit logic is used which is less efficient and may lead to errors \\ \hline
    S31 & ES & 227 & '==' or '!=' is being used to compare Strings instead of the equals() method \\ \hline
    S32 & VA & 8 & A primitive array is passed to a method with a variable number of arguments, resulting in the array being treated as a single argument, which may not be expected behaviour \\ \hline
    S33 & ICAST & 181 & A type cast of numeric values is performed in an incorrect or useless way \\ \hline
    S34 & DCN & 263 & A null check should be performed instead of catching a NullPointerException \\ \hline
    S35 & NN & 31 & notify() or notifyAll() is used without any modification to a mutable field \\ \hline
    S36 & RR & 58 & The return value of the read() method of an InputStream is ingored, potentially leading to sporadic failures \\ \hline
    S37 & ENV & 5 & It is preferable to use portable Java system properties instead of environment variables where possible \\ \hline
    S38 & UUF & 7194 & A field is never used \\ \hline
    S39 & PA & 785 & A public field should be made less visible \\ \hline
    S40 & SQL & 105 & An SQL call is made in an incorrect way, either allowing for SQL injection attacks, or incorrectly accessing results \\ \hline
    S41 & DE & 181 & An exception may be ignored instead of handled \\ \hline
    S42 & VO & 71 & A field or reference is treated as volatile when it may be preferable for it to be non-volatile \\ \hline
    S43 & DMI & 451 & A method is invoked in an inefficient, incorrect, or insecure way \\ \hline
    S44 & INT & 74 & A integer operation does not perform useful work, or a numeric value is compared to a value gauranteed to be outside its range \\ \hline
    S45 & CO & 47 & A compare() or compareTo() method is incorrectly implemented \\ \hline
    S46 & EQ & 262 & An equals() method is incorrectly implemented \\ \hline
    S47 & RE & 16 & A regex value is potentially either used incorrectly or used unintentionally \\ \hline
    S48 & HE & 87 & An unhashable class is used in a hashable context, or the class may violate the invariant that equal objects should have equal hash codes \\ \hline
    S49 & GC & 18 & An argument is used which has a type that is potentially not compatible with the expected generic parameter type. \\ \hline
    S50 & UW & 17 & A call to wait() is made without a guard condition \\ \hline
    S51 & WA & 26 & A call to wait() or await() is made outside a loop in a context that may have multiple conditions being observed \\ \hline
    S52 & DL & 6 & Synchronization is performed on an object that could be shared amongst all objects in the JVM, resulting in a potential deadlock \\ \hline
    S53 & LI & 80 & A static field is lazily initialized without synchronization, which could result in incorrect multi-thread behaviour \\ \hline
    S54 & DB & 35 & Two or more conditional branches or switch statements use the exact same code. \\ \hline
    S55 & IS & 44 & A guarded field is not properly guarded against concurrent access, or a field is accessed with inconsistent synchronization \\ \hline
    S56 & URF & 747 & A field seems to never be read \\ \hline
    S57 & BC & 194 & An impossible cast is made, or the cast is unchecked \\ \hline
    S58 & STCAL & 145 & A Calendar or DateFormat object is accessed in a way indicative of multithreading, even though these types are not threadsafe \\ \hline
    S59 & RC & 107 & Two references are compared using '==' or '!=' instead of the equals() method \\ \hline
    S60 & UR & 54 & A constructor performs a read of a value that has not yet been initialized \\ \hline
    S61 & WL & 3 & Synchronization is performed on the getClass() return value instead of a class literal, leading to potential data races amongst subclasses \\ \hline
    S62 & EC & 38 & An equality comparison is performed in a way that is likely to always result in the objects being inequal \\ \hline
    S63 & UPM & 90 & A private methods seems to never be called \\ \hline
    S64 & FE & 21 & Two floating point values are compared for equality in a way that could fail (due to potential rounding), or a floating point value is compared to the special NaN value (which will always be inequal) \\ \hline
    S65 & ODR & 259 & A database resource seems to not be closed on some execution paths \\ \hline
    S66 & BX & 991 & Boxing and/or unboxing is performed in an inefficient way \\ \hline
    S67 & IM & 81 & An operator may be used in an incorrect or unreliable way \\ \hline
    S68 & SC & 22 & A constructor starts a thread, which may behave incorrectly for subclasses \\ \hline
    S69 & IA & 23 & An inner class invokes a method in a way which could be ambiguous \\ \hline
    S70 & ME & 42 & A mutable enum field can be set from outside its package, and can be changed either accidentally or maliciously \\ \hline
    S71 & CN & 71 & A class' clone() method does not call super.clone(), or a class implementing Cloneable does not implement clone() or vice versa \\ \hline
    S72 & UCF & 23 & A control flow statement has no effect on the code's execution whether the branch is taken or not \\ \hline
    S73 & MWN & 2 & Object.wait(), Object.notify(), or Object.notifyAll() is called, without having an obvious lock on the object \\ \hline
    S74 & TLW & 2 & A wait is performed while having multiple locks, which may result in a deadlock \\ \hline
    S75 & HRS & 16 & An HTTP header or cookie is constructed in a way that could lead to a HTTP response splitting vulnerability \\ \hline
    S76 & IL & 33 & An infinite loop is created, either through a loop with no exit condition, unguarded recursive calls, or adding a collection to itself \\ \hline
    S77 & IT & 32 & A class implements the Iterator interface, however its next() method can not throw java.util.NoSuchElementException \\ \hline
    S78 & SA & 64 & A field is self-assigned, or an apparently useless self-computation or self-comparison is performed \\ \hline
    S79 & SS & 115 & An instance field appears to be a c compile-time static value, and should likely be a static field \\ \hline
    S80 & RU & 3 & Invokes run() on an object, where Thread.start() might be more appropriate \\ \hline
    S81 & AT & 42 & Calls to a concurrent abstraction may not be executed atomically \\ \hline
    S82 & FI & 17 & A finalizer is incorrectlly or inefficiently implemented \\ \hline
    S83 & MC & 5 & An overrideable method is called from a constructor or clone() method, which may result in the method being called while not fully initialized \\ \hline
    S84 & UWF & 104 & A field is never set, or is only ever set to null \\ \hline
    S85 & DP & 45 & Executed method call may require security permission, and should be executed inside a doPrivileged block \\ \hline
    S86 & FL & 16 & A floating point value is used in a way where its lack of precision may be detrimental \\ \hline
    S87 & RpC & 26 & A conditional test is performed more than once sequentially, which is useless \\ \hline
    S88 & JCIP & 14 & A non-final field is created in an immutable class \\ \hline
    S89 & RS & 1 & A serializable class' readObject() method is synchronized, even though it should only be accessible by one thread \\ \hline
    S90 & MF & 14 & A field is masked by a local variable or by a subclass' field, which could result in confusing or incorrect behaviour \\ \hline
    S91 & TQ & 1 & A type qualifier is either potentially missing, or potentially used incorrectly \\ \hline
    S92 & IP & 8 & A parameter value is ignored before being overwritten \\ \hline
    S93 & IMSE & 1 & An IllegalMonitoringStateException is caught, even though this exception is normally only thrown due to a code design flaw \\ \hline
    S94 & SWL & 8 & Thread.sleep() is called while a lock is held, leading to poor performance or a potential deadlock \\ \hline
    S95 & UI & 6 & A getResource() call is made that could behave unexpectedly if the calling class is extended \\ \hline
    S96 & OVERRIDING & 15 & Super method is annotated with @OverridingMethodsMustInvokeSuper, but the overriding method does not invoke it \\ \hline
    S97 & UMAC & 12 & An anonymous class defines a method which is not invoked, and seems to be otherwise uncallable \\ \hline
    S98 & XSS & 2 & This code potentially introduces a cross-site scripting vulnerability \\ \hline
    S99 & QF & 1 & A for loop's incrementation seems to be potentially incorrect  \\ \hline
    S100 & CNT & 12 & A constant seems to be approximately equal to a known library value (e.g. approximately equal to Math.PI) \\ \hline
    S101 & BSHIFT & 1 & Order of operations for a binary shift may be wrong here \\ \hline
    S102 & FS & 726 & A format string should use \%n instead of the newline character \\ \hline
    S103 & IC & 5 & A circular reference was detected in static initializers \\ \hline
    S104 & LG & 3 & A logger configuration may be lost due to an incompatibility in OpenJDK \\ \hline
    S105 & SP & 2 & The compiler may move a field read outside a loop, which could result in an infinite loop \\ \hline
    S106 & J2EE & 2 & A non-serializable object may be getting stored in an HttpSession, which could result in an error \\ \hline
    S107 & SR & 66 & The return value of the skip() method of an InputStream is ingored, potentially leading to sporadic failures \\ \hline
    S108 & VSC & 62 & A security check method is non-private and non-final, potentially leaving it vulnerable to being overridden \\

\label{table:issueDefs}
\end{longtable}

\end{appendices}

\end{document}